\documentclass[fleqn,usenatbib]{mnras}

\usepackage{newtxtext,newtxmath}
\usepackage{kotex}
\usepackage{CJKutf8}

\usepackage[T1]{fontenc}
\newcommand{\Mp}{M_{\rm{P}}}

\DeclareRobustCommand{\VAN}[3]{#2}
\let\VANthebibliography\thebibliography
\def\thebibliography{\DeclareRobustCommand{\VAN}[3]{##3}\VANthebibliography}

\DeclareRobustCommand{\DE}[3]{#2}
\let\DEthebibliography\thebibliography
\def\thebibliography{\DeclareRobustCommand{\DE}[3]{##3}\DEthebibliography}

\DeclareRobustCommand{\DI}[3]{#2}
\let\DIthebibliography\thebibliography
\def\thebibliography{\DeclareRobustCommand{\DI}[3]{##3}\DIthebibliography}

\DeclareRobustCommand{\DU}[3]{#2}
\let\DUthebibliography\thebibliography
\def\thebibliography{\DeclareRobustCommand{\DU}[3]{##3}\DUthebibliography}

\usepackage{graphicx}	
\usepackage{amsmath}	
\usepackage[normalem]{ulem}

\DeclareMathOperator\erf{erf}

\title{Constraining Cosmic Inflation with Observations: Prospects for 2030}

\author[B. Bahr-Kalus, D. Parkinson and R. Easther]{
Benedict Bahr-Kalus,$^{1}$\thanks{E-mail: benedictbahrkalus@kasi.re.kr} David Parkinson,$^{1,2}$\thanks{E-mail: davidparkinson@kasi.re.kr}
and Richard Easther,$^{3}$\thanks{E-mail: r.easther@auckland.ac.nz}
\\
$^{1}$Korea Astronomy and Space Science Institute, Yuseong-gu, Daedeok-daero 776, Daejeon 34055, Korea\\
$^{2}$University of Science and Technology, Daejeon 34113, Korea\\
$^{3}$Department of Physics, University of Auckland, Private Bag 92019, Auckland, New Zealand
}

\date{Accepted XXX. Received YYY; in original form ZZZ}

\pubyear{2022}

\begin{document}
\label{firstpage}
\pagerange{\pageref{firstpage}--\pageref{lastpage}}
\maketitle

\begin{abstract}
The ability to test and constrain theories of cosmic inflation will advance substantially over the next decade. Key data sources include cosmic microwave background (CMB)  measurements and observations of the distribution of matter at low-redshift from optical, near-infrared, and 21cm intensity surveys. A positive detection of a CMB B-mode consistent with a primordial stochastic gravitational wave background (SGWB) is widely viewed as a smoking gun for an inflationary phase. Still, a null result does not exclude inflation. However, in a significant class of inflationary scenarios, a low SGWB amplitude is correlated with a more significant running, $\alpha_s$, in the primordial density perturbations than is seen with the  simplest inflationary potentials. With this motivation, we forecast the precision with which the spectral index $n_{\rm{s}}$ and $\alpha_{\rm{s}}$ 
can be constrained by currently envisaged observations, including CMB  (Simons Observatory, CMB-S4 and \textit{LiteBIRD}),   optical/near infra-red   (DESI and SPHEREx), and  21cm intensity mapping  (Tianlai and CHIME) surveys. We identify optimal combinations of datasets for constraining the running and show that they may yield additional and informative constraints on the overall inflationary parameter space if the SGWB remains undetected. 
\end{abstract}




\section{Introduction}

Cosmic inflation is a compelling explanation for the origin of the initial perturbations that seed large-scale structure formation. It generically predicts an initial spectrum of adiabatic, Gaussian, super-horizon  density perturbations, which grow under gravity to generate the distribution of matter and radiation  observed in present times \citep{Starobinsky:1979ty,Guth:1980zm,Linde:1981mu,Guth:1982ec,Hawking:1982cz,Starobinsky:1983zz}. These predictions are tested by and are consistent with observations of the cosmic microwave background \citep[CMB;][]{Liddle:1992wi,WMAP:2003elm,Planck:2018jri} anisotropies at recombination.

Inflation makes an additional prediction: the existence of a scale-free primordial stochastic gravitational wave background (SGWB), sourced by quantum fluctuations of spacetime during the inflationary epoch itself \citep{WMAP:2003elm,BICEP2:2018kqh,Planck:2018vyg,Tristram:2020wbi,BICEP:2021xfz}. These {\em tensor modes\/} distort  the linear polarisation angle of the CMB, generating a spectrum of B-modes in the polarisation distribution on the sky. Gravitational waves undergo minimal  reprocessing, and detecting an inflationary SGWB would provide a remarkable window into the early universe. Moreover, in the simplest inflationary models (single-field, slow-roll), there is even a direct \textit{consistency relation} between the scalar (density) and tensor (gravitational wave) power spectra \citep{Liddle:1992wi}. 

Detecting tensor modes in the CMB would be a  strong indication that inflation did happen in the very early universe and would permit decisive tests of inflationary models motivated by specific ultrahigh-energy physics scenarios. 
The SGWB is commonly parametrised via the ratio of amplitudes of the scalar (density) and tensor spectra, denoted $A_s$ and $A_T$, respectively. By convention, this is  measured at the comoving wavenumber $k=0.05\mathrm{Mpc}^{-1}$ and denoted 
\begin{equation}
    \label{eqn:tensortoscalar}
    r = \frac{A_T}{A_s}\, .
\end{equation}
However, there is no clear expectation regarding the likely amplitude of the inflationary SGWB. In fact, in many  scenarios, the  SGWB is effectively undetectable by any conceivable experiment.

Up until now, much of the focus in both the observational and theoretical communities has been on the value of $r$, given that a scale-free, primordial stochastic gravitational wave background would provide strong evidence for the existence of an inflationary phase in the very early universe. 
In 2021 the best constraints on $r$ moved from $r<0.06$ \citep{BICEP2Planck}  to $r<0.036$ at 95\% confidence \citep{BICEP:2021xfz}, via a combination of data from the BICEP3/Keck array and \textit{Planck}. Several  CMB polarisation experiments are in active development, including the Simons Observatory \citep{Ade:2018sbj}, CMB-S4 \citep{cmb-s4-r} and \textit{LiteBIRD} \citep{LiteBird}, the last with the potential to measure  $r$ to a precision  of $\sigma(r) \sim 10^{-4}$. 

Algebraically simple potentials favour larger values of $r$ and such scenarios can be successfully embedded in candidate theories of high-energy physics (e.g. \cite{Silverstein:2008sg}). However,  there is no requirement prevents $r$ from being vanishingly small. A failure to detect  primordial gravitational waves will not falsify the overall scenario but will only rule out `large-field' slow-roll inflation, a category which includes many once-canonical models.  Ironically, the more tightly $r$  is constrained, the more plausible a very low value becomes, since the theoretical lower bounds on its value lie many orders of magnitude below the reach of any plausible experiment. 

In the slow-roll framework, observables yield information about the shape of the inflaton potential. This connection underpins the inflationary inverse problem, or  ``reconstruction''  \citep{Lidsey:1995np}. In particular, specifying values of $n_s$ and $r$ sets the first two terms in the Hubble Slow Roll expansion. This yields a prediction for the amount of accelerated expansion that will transpire between the  pivot scale leaving the horizon and inflation coming to an end.    \cite{Easther:2021cpj} show that  current constraints on $n_s$ and $r$ imply all such two-parameter models predict an inflationary phase that lasts too long to be consistent with other cosmological constraints.  

There are several resolutions to this situation within the context of the overall inflationary paradigm. For example, inflation may end with a sharp discontinuity (a ``cliff edge''), or the potential may be too  complex to be accurately represented by the analogue of a low-order Taylor expansion. However, arguably the simplest resolution is working at next-to-leading order in the systematic slow-roll expansion. The additional term this introduces maps roughly to the third derivative of the potential, and,  given an appropriate value,  it ends inflation without excessive expansion. 

This term contributes to the  running, or scale dependence of $n_s$,  $\alpha_{\rm{s}} = dn_{\rm{s}}/ d\ln k$, where $k$ is the comoving  wavenumber.  For most  simple inflationary scenarios  $ 10^{-4} \lesssim |\alpha_{\rm{s}}| \lesssim 10^{-3}$ \citep{Adshead:2010mc}. However, if $\alpha_{\rm{s}}$ is chosen so that the duration of inflation is consistent with current observational constraints, we typically need $\alpha_s \lesssim -10^{-3}$. Moreover, if the bound on $r$ tightens the likely magnitude of $\alpha_s$ will increase \citep{Easther:2021cpj}.

These values are far still smaller than those currently allowed by  observations, $\alpha_{\rm{s}} =  -0.0045 \pm 0.0067$  (at 68\% confidence) \citep{Planck_running}. However, these constraints are set to tighten substantially in the coming decade, thanks to  measurements of the primordial density power spectrum from  large-scale galaxy surveys.\footnote{We focus  on the two-point function; the detection of primordial non-Gaussianity by any near-term experiment would immediately rule out all the  inflationary models considered here. } These include the optical  Dark Energy Spectroscopic Instrument (DESI) galaxy redshift survey \citep{DESI} and the Spectro-Photometer for the History of the Universe, Epoch of Reionization and Ices Explorer (SPHEREx) galaxy survey \citep{Dore:2014cca}, and  the Canadian Hydrogen Intensity Mapping Experiment (CHIME) \citep{chime} and Tianlai \citep{Xu:2014bya} 21cm intensity mapping experiments in radio.\footnote{These are far from  the only surveys. Other funded optical or 21cm intensity mapping surveys that will produce cosmologically relevant results before 2030 include the Baryon acoustic oscillations from integrated neutral gas observations project \citep[BINGO;][]{Abdalla:2021nyj}, \textit{Euclid} \citep{Amendola:2016saw}, the Javalambre-Physics of the Accelerated Universe Astrophysical Survey \citep[J-PAS;][]{J-PAS:2014hgg}, the Hydrogen Intensity and Real-time Analysis eXperiment \citep[HIRAX;][]{HIRAX}, the Legacy Survey of Space and Time (LSST) at the Vera C. Rubin Observatory \citep{LSST:2008ijt}, the MeerKAT Large Area Synoptic Survey \citep[MeerKLASS;][]{MeerKLASS:2017vgf}, the Widefield ASKAP L-band Legacy All-sky Blind surveY \citep[WALLABY;][]{Koribalski:2020ngl}, the WHT Enhanced Area Velocity Explorer \citep[WEAVE;][]{WEAVE:2016rxg}, and the 4-metre Multi-Object Spectroscopic Telescope \citep[4MOST;][]{4MOST}. However, we expect DESI, SPHEREx and CHIME to be the best of their classes for the purposes considered here. As explained in the text, we leave consideration of Lyman-$\alpha$ forest surveys, such as the Hobby-Eberly Telescope Dark Energy Experiment \citep[HETDEX;][]{Gebhardt:2021vfo} for future work. }

In what follows, we assemble a self-consistent compendium of forecasts for the cosmological constraints on ``inflationary observables'' expected from experiments currently deployed or under development. Given the  genuine possibility that  the  inflationary SGWB is vanishingly tiny, we focus on the corner of parameter space in which $n_{\rm{s}}$ and $\alpha_{\rm{s}}$ are the principal inflationary observables. We shall see that the larger values of $|\alpha_{\rm{s}}|$ that are associated with very small $r$  may begin to be accessible in the coming decade, even if they will not be rigorously testable on this timescale. 

The dynamical system that governs single-field slow-roll inflation and its associated observables is well studied. However, the  quality of present-day observational data significantly changes our priors regarding the possible form of the potential. The broader goal of this paper is to assess the ability of future CMB and large-scale structure surveys to constrain inflationary models in light of our current knowledge of the primordial perturbations. As already noted, the tension between scenarios described by only the first two slow-roll parameters and the data can be resolved in several ways. However, the existence of a well-delineated set of scenarios with low $r$ and large $\alpha_s$ provides an additional target for cosmological experiments. 

The paper is structured as follows. In Section~\ref{sec:inflation}, we sketch the connections between the power spectra, the slow-roll expansion and the expectations for ``inflationary''  observables. 
Section~\ref{sec:fisher}  summarises the Fisher forecast approach and the details of the surveys we will be simulating. In Section~\ref{sec:results} we forecast constraints from the different future CMB experiments and large-scale structure surveys for  $r$, $n_s$ and $\alpha_{\rm{s}}$, including joint constraints, with a focus on the running $\alpha_{\textrm{s}}$. We conclude in Section~\ref{sec:discussion}.

\section{Inflation and the Running Index}
\label{sec:inflation}
Single-field, slow-roll inflation is the simplest implementation of the broader inflationary paradigm. Its dynamics are governed by a field $\phi$ with potential $V(\phi)$ minimally coupled to Einstein's gravity, which obeys the equations of motion 
\begin{align} 
&H^2 =\frac{1}{3 \Mp^2}\left( \frac{\dot\phi^2}{2} + V(\phi)\right) \, , \label{eq:hsqrd} \\  
&\dot{H} = -\frac{1}{2\Mp^2} \dot{\phi}^2\, , \label{eq:hdot} \\
&\ddot{\phi} + 3 H \dot \phi + \frac{dV}{d\phi} =0 \, .
\end{align}
As usual, $a$ is the scale factor, $H$ is the Hubble parameter, and $\Mp$ is the reduced Planck mass which, in natural units, is related to Newton's constant via $G = 1/\Mp^2$.

The primordial observables are expressed in terms of the slow-roll parameters, which can be defined in terms of the potential, 
\begin{equation}
\epsilon = \frac{\Mp^2}{2} \left(\frac{V'}{V}\right)^2 \, ,\quad \eta = \Mp^2 \frac{V''}{V}\, ,\quad \xi = \Mp^4 \frac{V'V'''}{V^2}\, ,
\end{equation}
where a dash denotes a derivative with respect to $\phi$. To lowest order the key observables are  
\begin{equation} \label{eq:nsv}
n_s = 1- 6\epsilon_0 + 2 \eta_0 \, ,\quad r= 16\epsilon_0 \, , \quad \alpha_s = -2\xi_0 \, ,
\end{equation}
where the index $0$ denotes the value of the parameter  when the pivot scale leaves the horizon.
A  consequence of these relationships is that any set of inflationary observables can be mapped to an inflationary trajectory: given values for  $n_s$, $\alpha_s$ and $r$, one can predict the amount of accelerated growth that will occur after the pivot scale leaves the horizon on the assumption that higher order terms are negligible. The  dynamical system that describes single-field inflation has long been understood \citep{Lidsey:1995np}, and the possible inflationary mechanisms it supports are conveniently summarized in the so-called zoo-plot  \citep{Dodelson:1997hr}. 

Today, observational constraints  significantly sharpen priors on $n_s$ and $r$ and the corresponding slow-roll parameters. Recalling that $r\propto \epsilon$ and $\epsilon \propto (V'/V)^2$ we can divide the  parameter space into three distinct regimes,  $|V'| \gtrsim |V''|$, $|V'| \sim |V''|$ and $|V'| \lesssim |V''|$, with $\Mp=1$ \citep{Easther:2021cpj}. The first possibility is increasingly close to being observationally untenable. Conversely, if the CMB experiments  now under construction do not detect an SGWB, it will follow that $|V''| \gtrsim 5|V'|$, leaving only the third alternative. Consequently, even if the coming decade does not see a detection of an SGWB  constraints on the overall inflationary parameter space will still advance both   quantitatively and qualitatively.

The relationship between the spectral index and the derivatives of the potential in Eq.~\ref{eq:nsv} can be viewed as a differential equation,  which can be solved to yield a potential which delivers any pair of $n_s$ and $r$ at the pivot. This leads to the so-called {\em universality classes} \citep{Mukhanov:2013tua,Roest:2013fha,Creminelli:2014nqa,CMB-S4:2016ple} of inflation and is one basis of a widely repeated claim that $n_s \sim 1/N$ and $\alpha_s \sim -1/N^2$, where $N$ is the remaining number of e-folds of inflation. 

The correlation between the running and the duration of inflation has been explored a number of times \citep{Malquarti:2003ia,Makarov:2005uh,Easther:2006tv,Adshead:2008vn,Adshead:2010mc}. In the potential expansion, the scale-dependence of the $\eta$ parameter is given by 
\begin{equation}
\frac{d \eta}{dN} \approx  {M_P^4}  \left[ \frac{V'}{V} \frac{V'''}{V} - \left(\frac{V'}{V}\right)^2 \frac{V''}{V} \right]  \, ,
\end{equation}
where we have made use of the generic single-field slow-roll result 
\begin{equation}
    \frac{dN}{d\phi} = \frac{1}{\Mp} \frac{1}{\sqrt{2\epsilon}} \, .
\end{equation}
If $\epsilon$ and thus $V'/V$ is small, $(V'/V)^2$ is very small, while the measured value of $n_s$ fixes $V''/V$. In these circumstances, the $V' V'''/V^2$ term cannot be considered irrelevant without an implicit tuning.  In this case,  the  dynamical system underlying the inflationary phase is more effectively treated in the Hubble slow roll hierarchy \citep{Liddle:1994dx,Hoffman:2000ue,
Kinney:2002qn}, which can be explored analytically in this low-$\epsilon$ limit  \citep{Adshead:2008vn}. 

The gradual tightening of constraints on $r$ and $n_s$ has reached the point that any inflationary models described by only the first two slow roll parameters lead to an inflationary phase that is inconsistent with limits on the amount of accelerated expansion that occurs after astrophysical perturbations leave the horizon.\footnote{This quantity is not precisely defined unless the post-inflationary thermalisation mechanism is  specified  \citep{Dodelson:2003vq,Liddle:2003as,Adshead:2010mc,Munoz:2014eqa}. However, an upper bound follows from assuming immediate thermalisation as inflation ends.} For these cases, we can compute the value of $V'''$ that will ensure that inflation terminates appropriately, which also sets the value of $\alpha_s$ at the pivot \citep{Easther:2021cpj}. Given these assumptions, $|\alpha_s|$ is larger than the  estimate of $1/(1-n_s)^2$ derived from universality class arguments. 

In doing so, we implicitly assume that $V'''/V$  is the ``last relevant term'' in the Taylor expansion of the potential or, more accurately, its Hubble Slow Roll analogue. We then constrain $\alpha_s$ by specifying an overall duration of the inflationary phase,   assuming that inflation ends via the breakdown of slow-roll \citep{Easther:2021cpj,Stein:2022cpk}.    If either of these conditions is violated, the arguments here do not constrain $\alpha_s$. However, given current observational constraints on the primordial spectra, this scenario represents a simple, distinctive and testable class of inflationary model and is thus of significant interest.  

\begin{figure}
    \centering
    \includegraphics[width=\columnwidth]{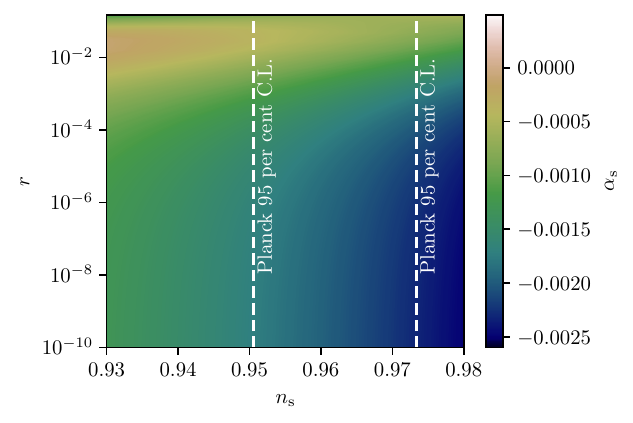}
    \caption{Contours of constant $\alpha_\mathrm{s}$ in the $n_\mathrm{s}$ − $\ln r$ plane with a
three-term slow-roll hierarchy and $\xi$ set to give $N = 55$ when
the pivot leaves the horizon. The vertical lines mark the current $95\%$ contours for $n_\mathrm{s}$ from \citet{Planck_running}.}
    \label{fig:contours}
\end{figure}

Fig.~\ref{fig:contours} illustrates the relationship between $n_s$, $r$  and  $\alpha_s$ if we assume that 55 e-folds of inflation occur after the pivot scale leaves the horizon. This implies rapid but not instantaneous thermalisation, and in scenarios with a long phase in which the post-inflationary universe is effectively matter-dominated (which is a distinct possibility), the corresponding value of $\alpha_s$ would increase.   While much attention is paid to the tightening bounds on $r$, the measured value of $n_s$ is also critical -- the closer this parameter is to unity, the larger $V'''$ and thus $\alpha_s$ must be to match constraints on the duration of inflation. Current bounds on $n_s$ are overlaid on Fig.~\ref{fig:contours}. If the permitted values of $n_s$ contract toward the centre of its current range as further data become available, expectations for the magnitude of $\alpha_s$ will likewise increase.

\section{Fisher forecasting}
\label{sec:fisher}

\subsection{Fisher matrix}

To make meaningful comparisons between multiple possible combinations of survey data we first need to create a mutually consistent set of forecasts. A forecast assumes some likelihood function $\mathcal{L}(\theta)$, where the probability of the data ($D$) can be evaluated with regards to assumed (fixed) values of the parameters of interest ($\theta$),
\begin{equation}
    \mathcal{L}(\theta) = P(D|\theta) \,.
\end{equation}
When faced with real data, this likelihood function can be used to evaluate the probability distribution of these parameters. But even before the data is taken, a forecast can be made to estimate the effectiveness of the planned survey or experiment. A common approach is the Fisher forecast \citep{KendallStuart,Tegmark1997,BMIC}, where the data is simulated assuming some true values of the parameters $\hat{\theta}$, and the log-likelihood function is Taylor-expanded around the extremum. For a given realisation of the future data, the expansion can be modelled as
\begin{equation}
    \ln \mathcal{L}(\theta) \approx \ln \mathcal{L}(\theta^*) + \frac12 \sum_{ij} (\theta_i-\theta^*_j)^T H_{ij} (\theta_i-\theta^*_j)\,,
\end{equation}
where $\theta^*$ is the maximum likelihood value of the parameters, and $H$ is the Hessian of the log-likelihood,
\begin{equation}
    H_{ij} = \left. \frac{\partial^2 \ln \mathcal{L}}{\partial \theta_i \partial \theta_j} \right\vert_{\theta^*}\,.
\end{equation}
Averaging over all possible realisations of the data, the maximum likelihood value $\theta^*$ should approach the fiducial value $\hat{\theta}$, and the Fisher matrix is the expectation of the Hessian
\begin{equation}
    F_{ij} = \langle H_{ij} \rangle\,.
\end{equation}
Thus the Fisher matrix encodes the curvature of the likelihood surface around the peak, and in the case of a Gaussian likelihood distribution, can be related to the parameter probability covariance matrix $\mathbf{C}$,
\begin{equation}
    \mathbf{F} = \mathbf{C}^{-1}\,.
\end{equation}
This is the limiting best case of the Cram\'er–Rao bound.

A forecast 68\% constraint on a parameter can therefore be obtained by computing the curvature of the likelihood with the respect to the parameters. In some cases, this can be performed analytically, but many likelihoods (and almost all of those we encounter in cosmology) have a non-linear relationship between input parameters and output value. Therefore the partial derivative will need to be computed through some finite differencing method.

Since the values of the observable data and errors on the data are assumed fixed given the fiducial values,  the Fisher matrix depends on the sensitivity of the theoretical prediction for the observable with respect to the parameter(s) of interest. For the next generation of cosmological surveys,  accurate predictions of  observables become difficult, due to the presence of systematic effects associated with how the data is taken. This is already the case with  CMB power spectrum measurements, where foreground dust or synchrotron emission can be confused with primordial CMB anisotropies. These foregrounds can be modelled using nuisance parameters $\theta_{\mathbf{n}}$. Since we do not know these noise parameters to infinite precision we have included them in our Fisher analysis, alongside the standard set of cosmological parameters $\theta_{c}$.  The nuisance parameters will differ between  different types of observations (i.e. CMB vs optical galaxy redshift surveys), so  confusion between a cosmological parameter and a nuisance parameter for a single experiment may be alleviated in combination.

\subsection{Fiducial parameters}

Our primary goal is to assess the sensitivity of future surveys to parameters relating to the power spectra of primordial scalar and tensor perturbations. These parameters usually are independent of (and thus relatively uncorrelated with) the cosmological parameters that control the evolution of the late-time universe. We assume a concordance Flat $\Lambda$CDM cosmology, where the dark energy is a cosmological constant with no time evolution, and the total density is equal to the critical density, i.e. 
\begin{equation}
    \Omega_\Lambda = 1 - \Omega_{\mathrm{CDM}} - \Omega_b\,.
\end{equation}
The primordial power spectrum for the scalar perturbations is parameterised as
\begin{equation}
    P_s(k) = A_s \left(\frac{k}{k_0}\right)^{n_s -1 + \frac12 \alpha_{\rm{s}}\ln(k/k_0) + \ldots} \, .
    \label{eq:scalar_pow}
\end{equation}
We assume that the running of the running $\beta_s=d\ln\alpha_{\rm{s}}/d\ln k=0$, and that higher order derivatives are also absent. We also assume the absence of any features (e.g. \citet{Adams:2001vc,2014JCAP...08..048H}) or cut-off (e.g. \citet{2014JCAP...08..053A}) in the power spectrum.

For the spectrum of tensor perturbations, the amplitude is expressed in terms of the tensor-to-scalar ratio (Eq. \ref{eqn:tensortoscalar}). We work with an inflationary prior so that the slow-roll consistency relation
\begin{equation}
    n_T = -\frac{r}{8}\,. 
\end{equation}
fixes the index of the tensor power spectrum.
The fiducial values of the  cosmological parameters are chosen to be the \textit{Planck} best fit cosmological parameters and  are listed in Tab.~\ref{tab:fiducialcosmo}.
\begin{table}
\centering
\begin{tabular}{l l c l} \hline
 Name & Symbol & Value & \\ \hline
 Physical baryon density & $\Omega_b h^2$ & 0.02237 & varying \\ 
 Physical CDM density & $\Omega_{\mathrm{CDM}} h^2$ & 0.12 & varying\\  
 Hubble parameter & $H_0$ & 67.36 & varying \\   
 Matter density & $\Omega_{\rm m}$ & $\Omega_b + \Omega_{\mathrm{CDM}}$ & derived\\
 Cosmological constant density &$\Omega_\Lambda$ & $1 - \Omega_{\rm m}$ & derived\\
 Optical depth to reionization &$\tau$ & 0.0925 & varying\\
 Physical neutrino density$^\dagger$ & $\Omega_{\nu} h^2$ & $4.28\times10^{-4}$ & fixed\\
 Curvature parameter & $\Omega_K$ & $0$ & fixed\\
 Dark energy equation of state & $w$ & $-1$ & fixed\\
 Primordial spectral index & $n_{\rm s}$ & 0.9649 & varying \\
 ~~of scalar fluctuations & & \\
 Amplitude of scalar fluctuations & $A_{\rm s}$  & $2.215\times10^{-9}$ & varying \\ 
 Running of primordial & $\alpha_{\rm s}$ & 0 & varying \\ ~~spectral index & & \\
 Tensor to scalar ratio & $r$ & 0 & varying \\\hline
\end{tabular}
\caption{Fiducial values assumed in our Fisher forecasts.\label{tab:fiducialcosmo}}
\end{table}

\subsection{Future surveys}

A number of cosmological surveys are currently in different phases of design and construction. To forecast how constraints on the primordial power spectrum will improve over the period 2022-2030, we focus on those that are currently deployed or an advanced state of preparation, namely

\begin{itemize}
    \item \textbf{SO:} Simons Observatory \citep{Ade:2018sbj} - a ground-based CMB experiment being built at the same site as the Simons Array and the Atacama Cosmology Telescope. SO consists of four telescopes; three small aperture telescopes (SATs) and one large aperture telescope (LAT). It will cover the wavelength range 27-280 GHz. 
    \item \textbf{CMB-S4} \citep{S4conceptdef}\textbf{:} a ground-based CMB experiment being constructed at both the South Pole and Atacama desert sites.
    \item \textbf{LiteBIRD:} Lite (Light) satellite for the studies of B-mode polarization and inflation from cosmic background Radiation Detection \citep{LiteBird} - a satellite CMB observatory covering the wavelength range 40 to 400 GHz.  
    \item \textbf{DESI:} Dark Energy Spectroscopic Instrument, \citep{DESI} - a ground-based optical spectroscopic galaxy redshift survey using the Mayall 4m telescope at Kitt Peak. It will measure the redshifts of Luminous Red Galaxies (LRGs) up to $z=1$, Emission-Line Galaxies (ELGs) up to $z=1.7$, and Quasars (QSOs) in the range $2.1 < z < 3.5$. The survey will cover $\approx 14,000~\mathrm{deg}^2$.
    \item \textbf{SPHEREx:} Spectro-Photometer for the History of the Universe, Epoch of Reionization, and Ices Explorer \citep{Dore:2014cca} is a space-based observatory that will conduct an all-sky survey in the near-infrared, with photometry over the range 0.75–5.0 microns. Redshifts are determined photometrically by fitting templates to the NIR spectra, in particular, the 1.6 $\mu$m bump. Its launch date is planned for April 2025.
    \item \textbf{CHIME:} Canadian Hydrogen Intensity Mapping Experiment \citep{chime} is a ground-based radio observatory that will detect neutral hydrogen from total 21-cm emission over some relatively de-localised angular region in the sky. It  will cover 60\% of the sky area over the redshift range $0.8 < z < 2.5$. 
    \item \textbf{Tianlai} \citep{Xu:2014bya}\textbf{:} a ground-based radio observatory, which will survey the distribution of neutral hydrogen in the universe, in a similar manner to CHIME.
\end{itemize}

\subsection{CMB Fisher forecasting}
\label{sec:cmbfisher}

\begin{table}
    \centering
    \begin{tabular}{l|c|c|c}
        \hline
        Field & $f_\mathrm{sky}$ & $\ell_\mathrm{min}$ & $\ell_\mathrm{max}$\\
        \hline
        \multicolumn{4}{l}{\bf Simons Observatory Large Aperture Telescope} \\
        - T, E & 0.4 & 30 & 8000\\
        - B & 0.4 & 30 & 1000\\
        \multicolumn{4}{l}{\bf Simons Observatory Small Aperture Telescope} \\
        - E & 0.1 & 30 & 8000\\
        - B & 0.1 & 30 & 300\\
        \multicolumn{4}{l}{\bf CMB-S4 Large Aperture Telescope} \\
        - T & 0.4 & 30 & 3000\\
        - E & 0.4 & 30 & 5000\\
        \multicolumn{4}{l}{\bf CMB-S4 Small Aperture Telescope} \\
        - E, B & 0.03 & 21 & 335 \\
        \multicolumn{4}{l}{\bf LiteBIRD} \\
        - T, E, B & 0.9 & 2 & 200 \\\hline
    \end{tabular}
    \caption{Observed sky fractions $f_\mathrm{sky}$ and multipole ranges expected for the CMB experiments considered in this work.}
    \label{tab:fsky_and_ells}
\end{table}

\begin{figure}
    \centering
    \includegraphics[width=\columnwidth]{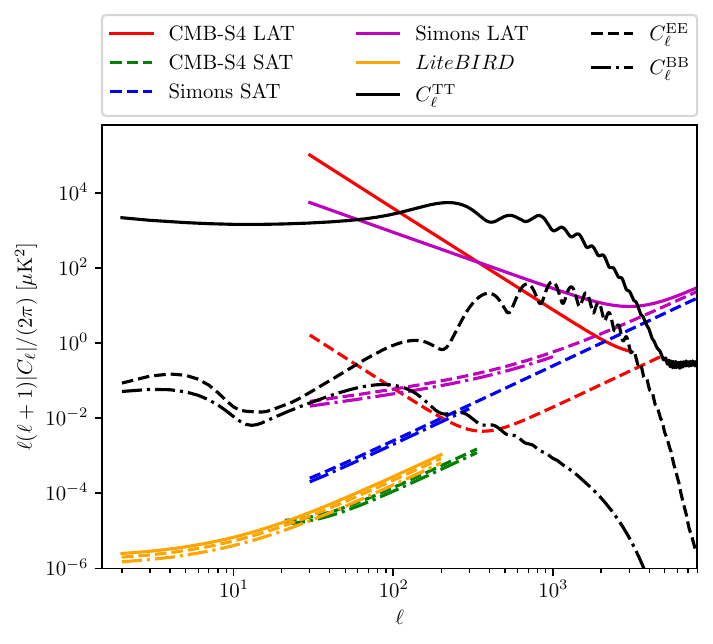}
    \caption{Angular power spectra (in black) and noise spectra (in colour) used in our CMB forecasts. The EE and BB noise spectra are slightly shifted towards the bottom in order to make them visible where they coincide with the temperature noise.}
    \label{fig:cmb_noise}
\end{figure}

CMB experiments observe large areas of the (spherical) sky so it is convenient to analyse their data using spherical harmonics $Y_\ell^m(\theta,\phi)$ such that the fluctuations $\delta X$ of the field $X\in \lbrace T, E, B\rbrace$ in direction $(\theta,\phi)$ can be written as
\begin{equation}
    \delta X(\theta,\phi) = \sum_{\ell m}a_{\ell m}^X Y_\ell^m(\theta,\phi).
\end{equation}
Assuming statistical isotropy and Gaussian fluctuations, we can fully describe the statistics of the fluctuation fields by their angular power spectra
\begin{equation}
    C^{XY}_\ell = \left\langle\left\vert a_{\ell m}^{X\ast} a_{\ell m}^Y\right\vert\right\rangle.
\end{equation}
Describing the likelihood function $\mathcal{L}$ in terms of angular power spectra $C^{XY}_\ell$, the Fisher matrix reads \citep[e.g.][]{Bermejo-Climent:2021jxf}
\begin{equation}
    F_{ij} = \frac{1}{2}\operatorname{tr}\left\lbrace \frac{\partial \mathbf{C}}{\partial \theta_i}\mathbf{C}^{-1}\frac{\partial \mathbf{C}}{\partial \theta_j}\mathbf{C}^{-1}\right\rbrace,
\end{equation}
where $\mathbf{C}^{XY}\equiv C^{XY}_\ell + N_\ell\delta_\mathrm{K}^{XY}$ and, to take the trace, one has to take into account that the number of modes corresponding to each multipole is given by $(2\ell + 1)f_\mathrm{sky}$. The observed sky fractions $f_\mathrm{sky}$ and the $\ell$-ranges considered when taking the trace are listed in Table~\ref{tab:fsky_and_ells}. We adopt the noise model 
\begin{equation}
    N_\ell = N_\mathrm{red}\left(\frac{\ell}{\ell_\mathrm{knee}}\right)^{\alpha_\mathrm{knee}} + N_\mathrm{white}
    \label{eq:CMBnoiseNl}
\end{equation}
from \citet{Ade:2018sbj} with the parameters listed in Table \ref{tab:CMBnoiseparameters}. The values of these parameters vary between frequency bands. By observing in multiple frequency bands, CMB experiments can distinguish cosmological signals from foreground contamination, whose spectral dependence is typically different from the blackbody curve of the CMB itself. A conservative assumption for the noise of the combined CMB maps is that it is equal to the minimum noise of each frequency band. We, therefore, choose effective noise parameters in a way that, for any given multipole $\ell$, Eq. \eqref{eq:CMBnoiseNl} approximates the minimum $N_\ell$ among all frequency bands.    The parameter values for the Simons Observatory are modelled after \citet{Ade:2018sbj}, the ones for CMB-S4 after \citet{S4conceptdef}, and we take $N_\mathrm{white}$ predictions for \textit{LiteBIRD} from \citet{Sailer:2021yzm}. The remaining \textit{LiteBIRD} noise parameters are taken from \citet{Nerval}. The noise models are plotted in Figure \ref{fig:cmb_noise}.

\begin{table}
    \centering
    \begin{tabular}{l|c|c|c|c}\hline
        Instrument & $N_\mathrm{white}$ [$\mathrm{\mu K^2}$]& $N_\mathrm{red}$ [$\mathrm{\mu K^2}$]& $\ell_\mathrm{knee}$ & $\alpha_\mathrm{knee}$\\\hline
        \multicolumn{5}{l}{\bf Simons Observatory Large Aperture Telescope} \\
        - temperature & $2.8\times 10^{-6}$ & $1.8\times 10^{-4}$ & 1000 & -3.5\\
        - polarisation & $2.8\times 10^{-6}$ & $2.8\times 10^{-6}$ & 700 & -1.4\\
        \multicolumn{5}{l}{\bf Simons Observatory Small Aperture Telescope} \\
        - polarisation & $3.1\times 10^{-7}$ & $3.1\times 10^{-7}$ & 25 & -2.5\\
        \multicolumn{5}{l}{\bf CMB-S4 Large Aperture Telescope} \\
        - temperature & $1.5\times 10^{-7}$ & $1.5\times 10^{-7}$ & 3400 & -4.7\\
        - polarisation & $1.5\times 10^{-7}$ & $1.5\times 10^{-7}$ & 340 & -4.7\\
        \multicolumn{5}{l}{\bf CMB-S4 Small Aperture Telescope} \\
        - polarisation & $8.5\times 10^{-8}$ & $8.5\times 10^{-8}$ & 60 & -2.5 \\
        \multicolumn{5}{l}{\bf LiteBIRD} \\
        - polarisation & $1.7\times 10^{-7}$ & $1.7\times 10^{-7}$ & 12 & -1.5 \\\hline
    \end{tabular}
    \caption{Effective noise-model parameters used in CMB forecasts.}
    \label{tab:CMBnoiseparameters}
\end{table}

A primordial B-mode signal is expected to be dominated by secondary B-modes generated by gravitational lensing. Consequently, `delensing' -- reconstructing the lensing potential and subtracting the lensing contribution to the B-modes -- will be key to the analysis of this data. In practice, this cancellation is likely to be incomplete, and  we introduce a nuisance parameter $A_\mathrm{lens}$ to marginalise over the residual.  \citet{Sherwin:2015baa} predict that half of the lensing contribution can be subtracted from the Simons Observatory B-mode signal. For CMB-S4, we use the conservative estimate  $A_\mathrm{lens} = 0.27$ where delensing is supported by the large area telescopes based in Chile \citep{S4delensingeff}.  

Galactic thermal dust and synchrotron radiation add significantly  to the polarisation  signal. The  spectra of these two effects are consistent with power laws whose exponents we take from \citet{Planck:2014dmk} for the thermal dust and from \citet{Choi:2015xha} for the synchrotron power spectrum. Including these two terms, our complete $C_\ell$ model reads
\begin{align}
    C_\ell^{XY} =& A_\mathrm{lens} C_\ell^{XY, \mathrm{lensed}} + \left(1 - A_\mathrm{lens}\right) C_\ell^{XY, \mathrm{raw}}\nonumber\\ 
    &+ A_\mathrm{dust}\ell^{-2.42} + A_\mathrm{synch}\ell^{-2.3} \text{(this line not for TT)},
\end{align}
where we obtain $C_\ell^{XY, \mathrm{lensed}}$ and $C_\ell^{XY, \mathrm{raw}}$ from the Cosmic Linear Anisotropy Solving System  \cite[CLASS]{Blas:2011rf,DiDio_Classgal}, and $A_\mathrm{dust}$ and $A_\mathrm{synch}$ are free parameters with zero-centred Gaussian priors with variance $5\times 10^{-4}$. This choice reproduces the forecasts presented in \citet{Ade:2018sbj}.

\subsection{LSS Fisher forecasting}
\label{sec:lss_fisher}
The most common summary statistic, when redshift information is available, is the Cartesian power spectrum
\begin{equation}
    P_x(\mathbf{k})=\frac{1}{(2\pi)^3}\left\langle \delta_x^\ast (\mathbf{k})\delta_x(\mathbf{k})\right\rangle
\end{equation}
of the over-density field 
\begin{equation}
    \delta_x(\mathbf{k}) = \int \delta_x(\mathbf{r})e^{i\mathbf{k}\cdot \mathbf{r}}\operatorname{d}^3\mathbf{r}
\end{equation}
of tracer $x$ in Fourier basis $\mathbf{k}$. However,  this is not a natural basis for dealing with wide-angle effects. For instance, ignoring the presence of a large magnification bias parameter $s_\mathrm{g}$ can render forecasts on $\sigma(n_\mathrm{s})$ too optimistic by a factor of 1.25 for Stage IV spectroscopic galaxy survey \citep{Bellomo:2020pnw}. Alternative approaches, such as the angular power spectrum in tomographic redshift bins \citep[e.g.][]{Asorey:2012rd,Loureiro:2018qva,Tanidis:2021uxp} or the 3D power spectrum in a spherical Fourier-Bessel basis \citep[e.g.][]{Heavens:2003jx,Leistedt:2011mk,Mike:2018zvb,Gebhardt:2021mds}, can naturally account for wide angle effects. However, the former loses some of the information contained in radial modes, while the latter complicates the cosmological interpretation at scales much smaller than the survey size. On the other hand, the 3D power spectrum has been widely used in the last decades, and many analysis tools exist. We, therefore, expect that the key science analyses in upcoming collaborations will still be done using Cartesian 3D power spectra, despite their shortcomings, and  make our forecasts accordingly.

\begin{figure}
    \centering
    \includegraphics[width = \columnwidth]{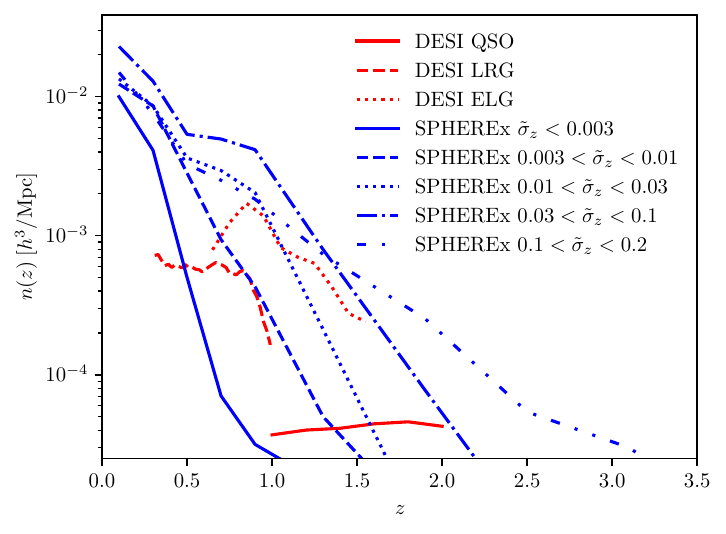}
    \caption{Assumed values for the number densities $n(z)$ used in DESI and SPHEREx forecasts.}
    \label{fig:n_of_z}
\end{figure}

Assuming a statistically isotropic universe, the angle-averaged power spectrum $P_\mathrm{g}(k)$ contains all information on inflation, even in redshift space. In this case, the Fisher matrix reads \citep{Tegmark:1997rp}
\begin{equation}
    F_{ij}=\frac{f_\mathrm{sky}}{2}\int_{k_\mathrm{min}}^{k_\mathrm{max}}\frac{\partial P_x(k,z)}{\partial \theta_i}\frac{\partial P_x(k,z)}{\partial \theta_j}\frac{V_k V_\mathrm{eff}(k, z)}{\left[P_x(k,z) + P_\mathrm{N}(z)\right]^2}\operatorname{d}k,
    \label{eq:PFisher}
\end{equation}
where $V_k=\frac{k^2}{(2\pi)^2}$ and $P_\mathrm{N}(z)$ is the noise power spectrum. If we use galaxies as tracers, the effective survey volume is \begin{equation}
    V_\mathrm{eff}(k, z) = V_\mathrm{survey}\left(\frac{\bar n(z) P(k, z)}{1 + \bar n(z) P(k, z)}\right)^2    
\end{equation}
for a survey with varying average density $\bar n(z)$ (such as galaxy surveys) or it is equal to the survey volume $V_\mathrm{survey}$ (e.g. for HI intensity mapping). We plot the values of $\bar n(z)$ that we assume for our forecasts in Figure \ref{fig:n_of_z}. We downloaded the SPHEREx $n(z)$ and $b(z)$ tables from \texttt{https://github.com/SPHEREx/Public-products}. The DESI values are inspired by preliminary target selection papers \citep{Raichoor:2020jcl,Yeche:2020tjm,Zhou:2020mgr}.

\begin{table}
    \centering
    \begin{tabular}{l|c|c|c|c|c}
        \hline
        Survey & $f_\mathrm{sky}$ & $k_\mathrm{min}$ & $k_\mathrm{max}$ & $b$\\
        \hline
        \multicolumn{5}{l}{\bf DESI} \\
        - QSO & 0.33 & 0.0016 & 0.24 & 2.4 \\
        - ELG & 0.33 & 0.0018 & 0.21 & 1.5 \\
        - LRG & 0.33 & 0.0025 & 0.16 & 2.4 \\
        \multicolumn{5}{l}{\bf SPHEREx} \\
        - $\sigma_z < 0.003 (1 + z)$ & 0.7 & 0.001 & 0.2 & 1.4 \\
        - $0.003 (1 + z) < \sigma_z < 0.01 (1 + z)$ & 0.7 & 0.001 & 0.2 & 1.2 \\
        - $0.01 (1 + z) < \sigma_z < 0.03 (1 + z)$ & 0.7 & 0.001 & 0.2 & 1.5 \\
        - $0.03 (1 + z) < \sigma_z < 0.1 (1 + z)$ & 0.7 & 0.001 & 0.2 & 1.4 \\
        - $0.1 (1 + z) < \sigma_z < 0.2 (1 + z)$ & 0.7 & 0.001 & 0.2 & 1.3 \\
        {\bf CHIME} & 0.6 & 0.0012 & 0.26 & 1.0 \\ \hline
    \end{tabular}
    \caption{Observed sky fractions $f_\mathrm{sky}$, $k$ ranges, and bias values $b$ expected for the LSS experiments considered in this work.}
    \label{tab:fsky_and_ks}
\end{table}

Another crucial ingredient for Equation \eqref{eq:PFisher} is the range of scales between $k_\mathrm{min}$ and $k_\mathrm{max}$ that each survey is going to be able to probe. For SPHEREx, we adopted $k_\mathrm{min} = 0.001$ and $k_\mathrm{max} = 0.2$ from \citet{Dore:2014cca}. For DESI and CHIME, we use $k_\mathrm{min} = 2\pi/\sqrt[3]{V_\mathrm{survey}}$ which we list in Table \ref{tab:fsky_and_ks}. This corresponds to the maximum scale attainable for a spherical survey with volume $V_\mathrm{survey}$. As neither DESI nor CHIME observe perfect spherical volumes, they will be able to measure longer modes in certain directions than those we consider here. However, these will be noisy and prone to direction-dependent systematic effects, and we do not take them into consideration. Since we only operate with a linear clustering model, we also do not consider scales above $k_\mathrm{max}$ where non-linear effects become important. We obtain $k_\mathrm{max}$ for DESI and CHIME by interpolating Table 1 of \citet{Seo:2003pu} for the mean redshift of the probe in question. With this choice of $k_\mathrm{max}$, we are only marginally affected by small-scale corrections from, e.g., massive neutrinos or alternative dark energy models.

\subsubsection{Galaxy Surveys}
Given a model matter power spectrum $P_\mathrm{m}(k)$ (that we can compute again using CLASS), we use the \citet{Kaiser:1987qv} formula to compute the model galaxy power spectrum 
\begin{equation}
    P_\mathrm{g}(k, z) = b^2P_\mathrm{m}(k,z)\left[1 + \frac{2\beta}{3} + \frac{\beta^2}{5}\right] .
    \label{eq:Kaiser}
\end{equation}
Here $b$ is the large-scale galaxy bias which we treat as a free parameter; $\beta$ is the linear growth rate of matter perturbations (which CLASS computes for a given set of cosmological parameters) divided by the galaxy bias $b$. We use this simple formula in our DESI forecasts where the spectroscopic redshift errors are negligibly small. For SPHEREx forecasts, however, redshift errors will be significant. The uncertainty in the redshift estimation of tracers washes out some of the information from radial modes.

Assuming Gaussian distributed redshifts with variance $\sigma_z$ and following the notation of \citet{Asorey:2012rd}, the 3D galaxy power spectrum reads
\begin{equation}
    P_\mathrm{g}(k, \mu, z) = b^2 \left(1+\beta\mu^2\right)^2 P_\mathrm{m}(k, z)e^{-k^2\sigma_\mathrm{t}^2(z)\mu^2},
\end{equation}
where $\mu$ is the cosine of the angle between the mode and the line-of-sight direction and $\sigma_\mathrm{t}(z)= {c\sigma_z(z)}/{H_\mathrm{fid}(z)}$. Note that we use the fiducial value $H_\mathrm{fid}(z)$ of the Hubble-Lema\^itre parameter since using the $H(z)$ of the varied cosmology would mean that the redshift uncertainty of the power spectrum would provide us with information on the expansion rate. Averaging over all angles $\mu$, we obtain the monopole power spectrum
\begin{align}
    \label{eq:PgIntegralresult}
    P_\mathrm{g}(k, z) = & b^2P_\mathrm{m}(k,z)\left[\sqrt{\pi}\left(4k^4\sigma_\mathrm{t}^4 + 4k^2\sigma_\mathrm{t}^2\beta + 3\beta^2\right)\erf(k\sigma_\mathrm{t}) \right.\nonumber\\
    & \left.- 2 k\sigma_\mathrm{t}e^{-k^2\sigma_\mathrm{t}^2}\beta\left(2k^2\sigma_\mathrm{t}^2(\beta+2)+3\beta\right)\right]/\left(8k^5\sigma_\mathrm{t}^5\right).
\end{align}

At scales where $k\ll\frac{1}{\sigma_\mathrm{t}}$, we can expand equation~\eqref{eq:PgIntegralresult} to find the well-known Kaiser formula (cf. Eq. \ref{eq:Kaiser}) with a quadratic correction in $k\sigma_\mathrm{t}$:
\begin{equation}
    P_\mathrm{g}(k, z) = b^2P_\mathrm{m}(k,z)\left[1 + \frac{2\beta}{3} + \frac{\beta^2}{5} - k^2\sigma_\mathrm{t}^2 \left(\frac{1}{3} + \frac{2\beta}{5} + \frac{\beta^2}{7}\right)\right].
\end{equation}
The shot noise of galaxies is already accounted for in $V_\mathrm{eff}(k, z)$. Thus, $P_\mathrm{N}(z) = 0$ in Equation \ref{eq:PFisher}.

\subsubsection{HI Intensity Mapping} 
When making forecasts for HI intensity mapping experiments, we proceed similarly as for spectroscopic galaxy surveys. Since intensity mapping cannot resolve single objects we model the power spectrum of the line intensity
\begin{equation}
    P_\mathrm{HI}(k,z) = \bar T_\mathrm{b}^2(z)b_\mathrm{HI}^2(z)P_\mathrm{m}(k,z)
\end{equation}
in terms of the temperature \citep{Santos:2015gra}
\begin{equation}
    T_\mathrm{b}(z)\approx 180h\frac{H_0}{H(z)}\Omega_\mathrm{HI}(z)(1+z)^2 \mathrm{mK},
\end{equation}
where we approximate the density of HI tracers
\begin{equation}
    \Omega_\mathrm{HI}(z) = 0.00048 + 0.00039z - 0.000065z^2
\end{equation}
with the fit of \citet{Cunnington:2019lvb}. As in the spectroscopic case, we treat the bias $b_\mathrm{HI}$ a free parameter. Following \citet*{Xu:2014bya}, we model the noise power spectrum as
\begin{equation}
    P_\mathrm{N}(k,z) = \frac{4\pi f_\mathrm{sky}\lambda^2 T_\mathrm{sys}^2 y(z) r_\mathrm{a}^2(z)}{A_\mathrm{e}\Omega_\mathrm{FoV} t_\mathrm{tot}},
\end{equation}
where $A_\mathrm{e}$ is the effective collecting area of each element, $\lambda$ is the redshifted wavelength of HI emission, $T_\mathrm{sys}$ designates the system temperature, $t_\mathrm{tot}$ the total integration time, $\Omega_\mathrm{FOV}$ symbolises the field of view of a single pointing, $y(z)=\frac{\lambda(1+z)}{H(z)}$ and $r_\mathrm{a}$ denotes the angular diameter distance.

\begin{figure*}
    \centering
    \includegraphics[width=1.4\columnwidth]{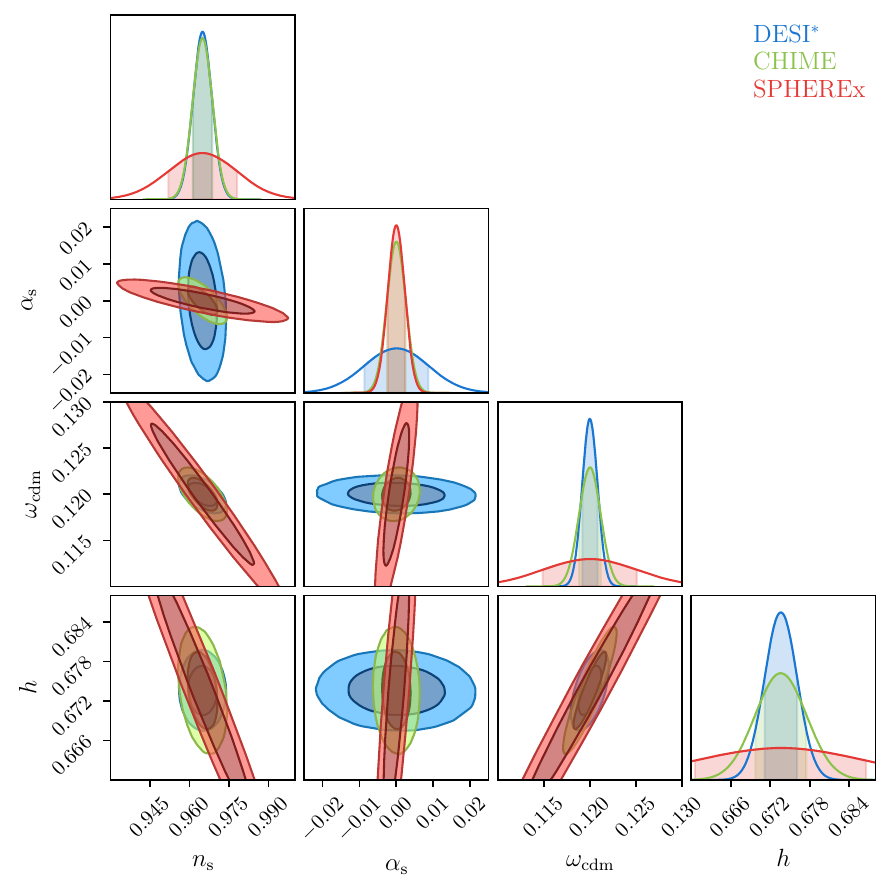}
    \caption{Forecast error contours on $n_\mathrm{s}$ and $\alpha_\mathrm{s}$ from LSS surveys. An asterisk in the label denotes that a Planck18 prior on the 6 \textit{Planck}-$\Lambda$CDM-model parameters has been applied. We do not apply any prior on $\alpha_\mathrm{s}$.}
    \label{fig:all_lss}
\end{figure*}

\begin{figure*}
    \centering
    \includegraphics[width=\textwidth]{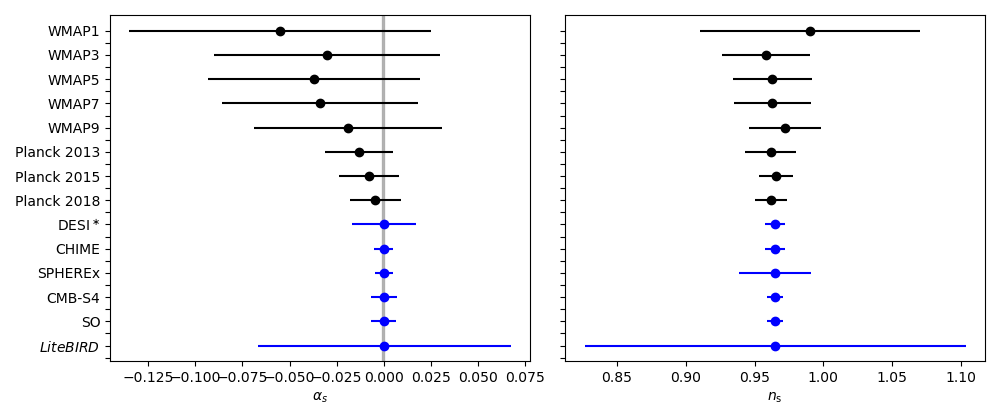}
   \caption{95\% confidence intervals for $\alpha_\mathrm{s}$ (left) and  $n_\mathrm{s}$ (right) from \textit{WMAP} and \textit{Planck} (in black) along with forecast constraints for upcoming surveys (in blue). The vertical area shaded in grey marks the $\alpha_\mathrm{s}$-range between $-10^{-3}$ and $-10^{-4}$. DESI${}^\ast$ refers to DESI forecasts with a \textit{Planck} 2018 prior on the 6 \textit{Planck}-$\Lambda$CDM-model parameters without a prior on $\alpha_\mathrm{s}$.}
    \label{fig:alpha_constraints}
    \label{fig:ns_constraints}
\end{figure*}

\section{Results}
\label{sec:results}
\subsection{CMB Forecasts}

\begin{figure}
    \centering
    \includegraphics[width=1.\columnwidth]{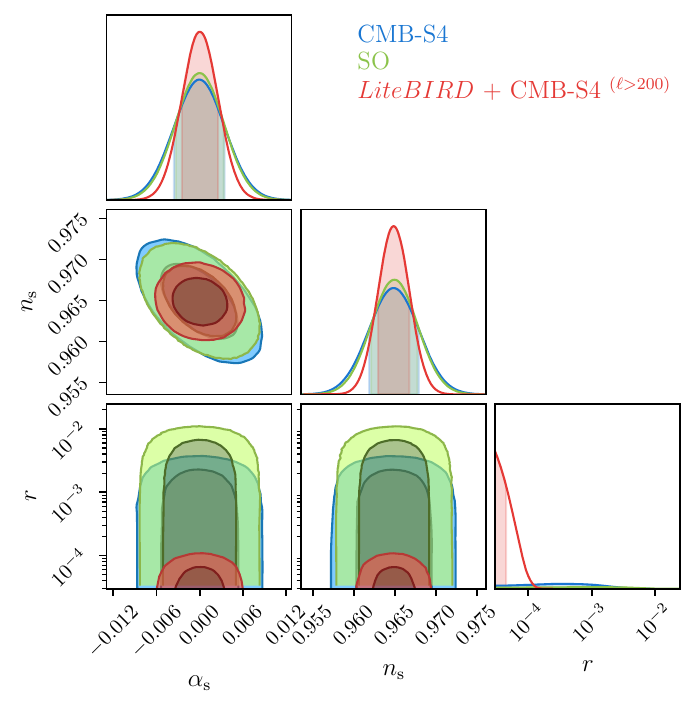}
    \caption{Forecast posterior contours of the spectral index $n_\mathrm{s}$, its running $\alpha_\mathrm{s}$ and the tensor-to-scalar ratio $r$ of the three CMB experiments considered in this work.}
    \label{fig:CMB_forecasts}
\end{figure}

\begin{table}
    \centering
    \begin{tabular}{l|c|c|c}
    \hline
       survey & $10^3 \sigma_{n_\mathrm{s}}$ & $10^3 \sigma_{\alpha_\mathrm{s}}$ & $10^4 r<$ \\\hline
       DESI, Planck18\footnote{The Planck18 prior here is just a prior on the 6-parameter $\Lambda$CDM model, i.e., there is no prior on $\alpha_\mathrm{s}$ and no measurement on $r$ presented here.} & 7.2 & 17 & - \\
       CHIME & 7.4 & 5.1 & - \\
       SPHEREx & 26 & 4.6 & - \\
       SO & 5.6 & ${}^{+6.5}_{-6.6}$ & 99 \\
       ~~ + DESI & 4.5 & 6.2 & 99\\
       ~~ + CHIME & ${}^{+2.9}_{-2.8}$ & 3.5 & 99 \\
       ~~ + SPHEREx & 1.6 & 2.4 & 98 \\
       S4 & 5.9 & ${}^{+6.8}_{-6.9}$ & 34 \\
       ~~ + DESI & 4.8 & 6.3 & 34\\
       ~~ + CHIME & 3.0 & 3.7  & 34 \\
       ~~ + SPHEREx & 1.7 & 2.4 & 33 \\
       ~~ + CHIME, DESI, SPHEREx & 1.5 & 1.8 & 33\\
       \textit{LiteBIRD} & 140 & 67 & 0.98\\
       \textit{LiteBIRD}, S4${}^{(\ell>200)}$ & 3.7 & 4.9 & 0.98\\
       ~~ + CHIME, SPHEREx & 1.2 & 1.7 & 0.97\\
       ~~ + CHIME, DESI, SPHEREx & 1.2 & 1.7 & 0.97\\ \hline
    \end{tabular}
    \caption{Forecast 95\% confidence intervals of the spectral index $n_\mathrm{s}$ and its running $\alpha_\mathrm{s}$, as well as forecast upper limits on $r$ at 95\% confidence level.}
    \label{tab:forecastlist}
\end{table}

We show forecasts of $n_\mathrm{s}$, $\alpha_\mathrm{s}$ and $r$ in Figure \ref{fig:CMB_forecasts} and list our expected constraints in Table \ref{tab:forecastlist}. The three CMB experiments under consideration, the Simons Observatory, CMB-S4 and \textit{LiteBIRD}, are complementary to each other. To measure the running $\alpha_\mathrm{s}$, we need a long lever arm measuring the spectral index on many different scales. From the CMB, we get those with the Large Aperture Telescopes (LAT) of the partial-sky, high-resolution experiments, Simons Observatory and CMB-S4. Interestingly, we shall see tighter constraints on $n_\mathrm{s}$ and $\alpha_\mathrm{s}$ from the Simons Observatory even though it is expected to run before CMB-S4. This behaviour is because the Simons LAT is less noisy at large scales than CMB-S4 LAT. However, CMB-S4 will greatly improve constraints on $r$ compared to the Simons Observatory, so the two experiments are complementary to each other.

While the partial-sky, high-resolution experiments will significantly tighten constraints on the running $\alpha_\mathrm{s}$, \textit{LiteBIRD} will provide strong constraints on $r$. Note that our forecasts depend, to some degree, on the choice of the pivot scale $k_0$ (cf. Eq. \ref{eq:scalar_pow}). We adopt $k_0 = 0.05\;\mathrm{Mpc}^{-1}$ throughout this work for internal consistency. \textit{LiteBIRD} is sensitive up to $\ell_\mathrm{max} = 200$ only, which roughly corresponds to $k\approx 0.0014\;\mathrm{Mpc}^{-1}$ in our fiducial cosmology. This scale is far below the pivot scale chosen here, and other authors commonly use $k_0 = 0.002\;\mathrm{Mpc}^{-1}$ in their \textit{LiteBIRD} forecasts. Using a larger pivot scale (i.e. smaller value of $k_0$), the \textit{LiteBIRD} constraints on $n_\mathrm{s}$ become tighter, though not competitive with  the Simons Observatory or CMB-S4. 

We find that the best we can do in terms of $n_\mathrm{s}$, $\alpha_\mathrm{s}$ and $r$ with CMB experiments in 2030 is to combine \textit{LiteBIRD} data at large scales ($\ell < 200$) and CMB-S4 at small scales ($\ell > 200$). Even though the Simons Observatory alone constrains $n_\mathrm{s}$ and $\alpha_\mathrm{s}$ better than CMB-S4 when combined with \textit{LiteBIRD}, \textit{LiteBIRD} beats both the Simons and CMB-S4 large-scale noise. At small scales, CMB-S4 outshines the Simons Observatory. Our forecast errors are $\sigma_{n_\mathrm{s}} = 0.0037$, $\sigma_{\alpha_\mathrm{s}} = 0.0049$ and $r < 9.8\times 10^{-5}$, assuming non-detection of positive $r$.

\subsection{LSS Forecasts}

Present constraints on $n_\mathrm{s}$ and $\alpha_\mathrm{s}$ have primarily come from the CMB since large-scale structure surveys have not covered large enough volumes yet to 
 measure the galaxy power spectrum at wavenumbers $k$  larger than the matter-radiation equality scale.  However, as galaxy surveys  probe larger volumes it will soon be feasible to measure $\alpha_\mathrm{s}$ and $n_\mathrm{s}$ from LSS surveys alone. Figure \ref{fig:all_lss} shows our forecast probability contours. As the position of the radiation equality scale depends on the Hubble-Lema\^itre rate $H_0$ and the density of cold dark matter $\omega_\mathrm{cdm}$, $n_\mathrm{s}$ and $\alpha_\mathrm{s}$ constraints from LSS are also dependent on these two parameters. 
 
The baryon acoustic oscillation (BAO) feature provides stringent constraints on $H(z)r_\mathrm{d}$ \citep{Bautista:2020ahg,deMattia:2020fkb,duMasdesBourboux:2020pck,Gil-Marin:2020bct,Hou:2020rse,Neveux:2020voa,Raichoor:2020vio}. Thus, to measure the Hubble-Lema\^itre rate $H_0$, one needs constraints on the sound horizon scale at the end of the Compton drag epoch $r_\mathrm{d}$ from CMB experiments. It is, therefore, customary to present LSS forecasts on cosmological parameters with a CMB prior. We show DESI contours with a Planck18 prior \citep{Planck:2018vyg} and refer to it as DESI$^\ast$ in the remainder of this work. We summarise all single survey constraints on $\alpha_\mathrm{s}$ in Figure \ref{fig:alpha_constraints} along with the 95\% confidence ranges from \textit{Planck} and \textit{WMAP} legacy data. Even though DESI alone will not provide tight constraints on $\alpha_\mathrm{s}$, with a Planck18 prior, it will provide an important semi-independent consistency check of the Planck18 measurement of $\alpha_{\rm{s}} =  -0.0045 \pm 0.0067$ \citep{Planck_running}. This is because the Planck18 prior does not put any limits on $\alpha_\mathrm{s}$ but only on the parameters $\omega_\mathrm{b}$, $\omega_\mathrm{cdm}$, $h$, $n_\mathrm{s}$, $A_\mathrm{s}$ and $\tau$. Here, we consider only direct-tracer measurements of DESI. However, DESI will also infer the matter distribution through Lyman-$\alpha$ (Ly-$\alpha$) forest features in the spectra of high redshift ($2.1 < z < 3.5$) quasars \citep{DESI:2022xcl}.
As the linear regime spans a wide range of $k$, Ly-$\alpha$ forest observations are, in principle, an excellent probe of $\alpha_\mathrm{s}$. \citet{DESI} forecast that constraints on $\alpha_\mathrm{s}$ from direct tracers will improve by almost a factor of three when combining them with Ly-$\alpha$ measurements. Even though this is an impressive improvement, we keep Ly-$\alpha$ forecasts for future work as this requires excellent control of broad-band systematics whose modelling requires more careful analysis \citep[for instance, along the lines of][]{Cuceu:2022brl}, and CHIME and SPHEREx are expected to provide similar $\alpha_\mathrm{s}$ constraints.
 
 In Figure \ref{fig:ns_constraints}, we show all single survey constraints on $\alpha_\mathrm{s}$ and $n_\mathrm{s}$. One can see that DESI* will tighten the Planck18 $n_{\textrm{s}}$ constraints on  $n_\mathrm{s}$. However, these will be beaten by the Simons Observatory. CHIME and SPHEREx are expected to provide tighter constraints on $\alpha_\mathrm{s}$ than CMB experiments but all the CMB experiments will  measure $n_\mathrm{s}$ more accurately than LSS surveys. While DESI does not quite cover enough volume to provide competitive $n_\mathrm{s}$ constraints, SPHEREx will not resolve small scales well enough to measure $\omega_\mathrm{cdm}$ with high accuracy, which is somewhat degenerate with $n_\mathrm{s}$ at large scales. 

\subsection{Combining CMB and LSS measurements}
\label{sec:combo}

\begin{figure}
    \centering
    \includegraphics[width=\columnwidth]{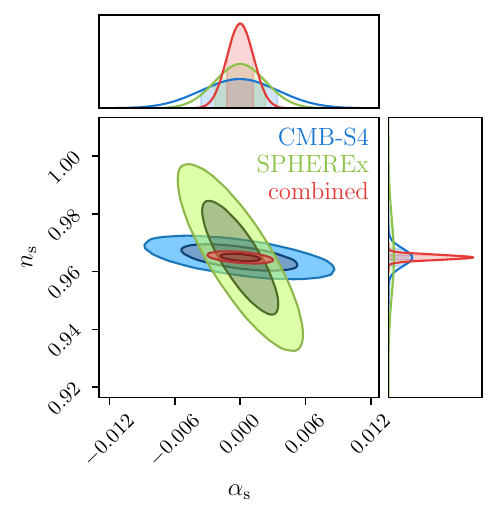}
    \caption{Forecast CMB-S4 and SPHEREx posterior contours as an example of how combining CMB data with large-scale structure data improves constraints on inflation.}
    \label{fig:S4_SPHEREx}
\end{figure}

\begin{figure*}
    \centering
    \includegraphics[width=\textwidth]{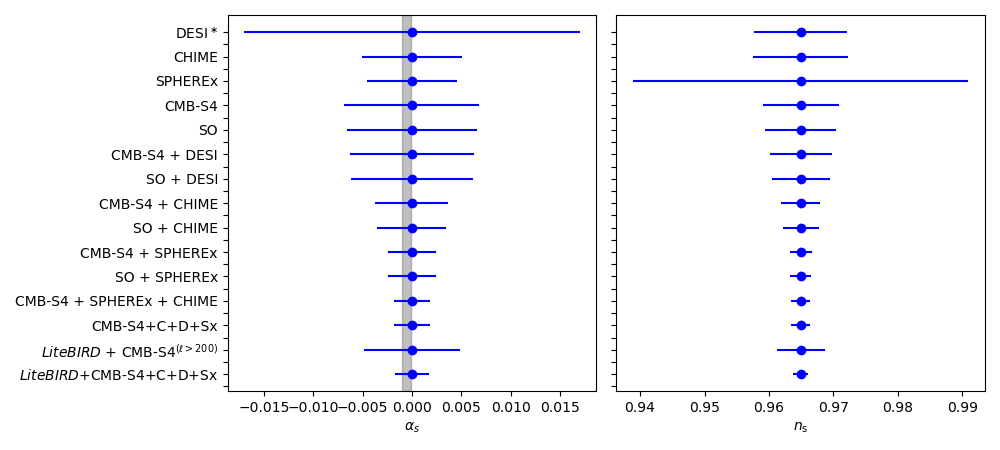}
    \caption{Similar to Figure \ref{fig:alpha_constraints} showing only forecasts including forecasts on data combinations. The vertical area shaded in grey in the left plot, again, marks the $\alpha_\mathrm{s}$-range between $-10^{-3}$ and $-10^{-4}$.}
    \label{fig:ns_constraints_forecasts_only}
\end{figure*}

As already discussed,  CMB experiments generally have a better handle on $n_\mathrm{s}$ than their LSS counterparts. On the other hand, by providing 3D information, LSS surveys  probe $n_\mathrm{s}$ on different scales, giving them more leverage on the running of the spectral index $\alpha_\mathrm{s}$. Assuming there are no tensions between CMB and LSS results we can  use this synergy to obtain tighter constraints on inflationary parameters. We illustrate this  in \autoref{fig:S4_SPHEREx} which shows the forecast probability contours in the $n_\mathrm{s}$-$\alpha_\mathrm{s}$ plane  for CMB-S4 and SPHEREx. CMB-S4 will yield tight error bars on $n_\mathrm{s}$ but will leave $\alpha_\mathrm{s}$ relatively unconstrained. With SPHEREx $n_\mathrm{s}$ and $\alpha_\mathrm{s}$ are perfectly degenerate, and we will not get any competitive constraints on either parameter. When combined with CMB-S4 data this degeneracy is broken, almost halving the CMB-S4 $n_\mathrm{s}$ and SPHEREx $\alpha_\mathrm{s}$ credible intervals.

\autoref{tab:forecastlist} lists  forecasts for all combinations of CMB and LSS probes that we considered. Even though DESI will add to  cosmological constraints when combined with Planck18, it will not significantly add to our knowledge about inflation once we have data from the Simons Observatory and CMB-S4. However, in the real world, where tensions can arise, DESI will provide key consistency checks on the new CMB data.

Combining CMB data with CHIME, SPHEREx, or both, we can learn substantially more about inflation than we can from these experiments on their own, as illustrated in the whisker plots of Figure~\ref{fig:ns_constraints_forecasts_only}. Merging CMB-S4 with data from DESI, CHIME and SPHEREx, we can hope to attain $\sigma_{\alpha_\mathrm{s}} = 0.0018$ at 95\% C.L. by 2030 or shortly thereafter. Consequently,  the class of scenarios with low $r$ and running $\alpha_\mathrm{s} \lesssim -10^{-3}$ identified in \cite{Easther:2021cpj} will be at the threshold of testability within the coming decade. 

\section{Conclusions and Discussion}
\label{sec:discussion}

The simplest implementation of inflation, i.e. two-parameter single-field slow-roll, has come under stress thanks to recent data from the BICEP3/Keck array and \textit{Planck}. The resulting bounds on $n_\mathrm{s}$ and $r$ suggest that  inflation would last too long \citep{BICEP:2021xfz,Easther:2021cpj} in these models to be consistent with broader consistency conditions on the duration of accelerated expansion. Including a non-trivial third derivative in the inflaton potential introduces an additional slow-roll parameter $\xi$ and, hence, a running spectral index $\alpha_\mathrm{s}$. By mapping the slow-roll parameters $\epsilon$, $\eta$ and $\xi$ to an inflationary trajectory, we can choose $\xi$ so that inflation has a physically reasonable  duration. To end inflation at $N = 55$ in this scenario requires $-0.0023 < \alpha_\mathrm{s} < -0.0004$ for any value of $n_\mathrm{s}$ within the \textit{Planck}-95\% range.

Taking a broader view, this analysis underlines the progress that has been made in observational cosmology since the first observation of primordial anisotropies in the CMB \citep{COBE:1992syq}.  In 1997, the inflationary parameter space included models with spectral indices on either side of unity \citep{Dodelson:1997hr} and large values of $r$. By the end of the 9-year {\it WMAP\/} campaign \citep{WMAP:2003elm,WMAP:2012nax} there was 5-$\sigma$ evidence that $n_{\mathrm{s}}$ was less than unity. These bounds tightened further with the advent of {\it Planck}, and the fraction of the $n_s$-$r$ plane in the original ``zoo plot'' that is consistent with observations has shrunk dramatically.

Looking to the future, we forecast the ability of the Simons Observatory, CMB-S4, \textit{LiteBIRD}, DESI, SPHEREx, and CHIME to constrain $r$, $n_\mathrm{s}$ and $\alpha_\mathrm{s}$ using the Fisher approach, assuming a \textit{Planck} six-parameter $\Lambda$CDM cosmology and a negligible primordial stochastic gravitational wave background. The CMB experiments, Simons Observatory, CMB-S4 and \textit{LiteBIRD}, are expected to tighten constraints on $n_\mathrm{s}$ and $r$ and, if inflation does not produce sizeable tensor perturbations, \textit{LiteBIRD} can determine that  $r < 10^{-4}$ (at 95\% C.L.). Such a bound on $r$ would provide further evidence for the claim that any inflaton potential contains higher derivatives. On the LSS side, due to its longer lever arms, we see tighter constraints on $\alpha_\mathrm{s}$, although these are degenerate with $n_\mathrm{s}$ within single surveys. For instance, we expect $\sigma_{\alpha_\mathrm{s}} \sim 0.005$ (again at 95\% C.L.) from CHIME and SPHEREx. Constraints from DESI are expected to be more than three times wider when inferring the matter density field from direct tracers only. 

Combining these constraints with estimates from the Lyman-$\alpha$ forest data from DESI may bring the error down to a similar level as with CHIME and SPHEREx. However, this would require a more careful analysis which we leave for future work. CMB experiments break the degeneracy in the LSS constraints, and the error on $\alpha_\mathrm{s}$ shrinks to 0.0017 using a combination of \textit{LiteBIRD}, CMB-S4 at $\ell > 200$, CHIME and SPHEREx. Thus, if inflation is sourced by a potential with a nontrivial third derivative, evidence for the running of the spectral index $\alpha_\mathrm{s}$ may be available by the end of this decade.

We have seen that we can expect two orders of magnitude improvement in measurements of $r$ and an order of magnitude improvement in measurements of $\alpha_{\mathrm{s}}$ over the coming decade. However, it is entirely possible that the number of non-trivial parameters in the ``inflationary sector'' of the concordance cosmology \citep{CMBPolStudyTeam:2008rgp} need not change at all between 1992 and 2032.  

That said, while there is no compelling argument as to why $r$ should be in the observable range, $\alpha_{\mathrm{s}}$ is non-zero in most-inflationary models.  \cite{Adshead:2010mc} catalogued $\alpha_{\mathrm{s}}$ in a variety of simple models, finding values that are typically ${\cal{O}}(10^{-4})$, and thus  several times smaller than the precision we expect from the coming generation of experiments. 

The scenario explored by \cite{Easther:2021cpj} provides a clear class of models in which the magnitude of  $\alpha_{\mathrm{s}}$ exceeds this generic expectation, and is close to the expected sensitivity of forthcoming experiments. This  provides an additional route by which next-generation experiments may usefully constrain a significant class of inflationary models.  As previously explained, a more considerable running is not {\em required\/} given that other mechanisms within the overall inflationary paradigm can limit the duration of inflation.  However, it is a largely unrecognised correlation and one that has become more interesting in the light of present data.  

To our knowledge, this work gives the first ``unified'' forecast with like-for-like comparisons for a wide variety of forthcoming experiments  and their possible combinations. We plan to extend this framework  to include more speculative scenarios with the longer-term goal of assisting the design of the successor generation of surveys by understanding how these proposals can interact to gain the tightest possible constraints on the overall inflationary parameter space.

\section*{Acknowledgements}
The authors would like to thank Cullan Howlett, Changbom Park, and Arman Shafieloo for valuable discussions around the project. BBK and DP are supported by the project \begin{CJK}{UTF8}{mj}우주거대구조를 이용한 암흑우주 연구\end{CJK} (``Understanding Dark Universe Using Large Scale Structure of the Universe''), funded by the Ministry of Science. RE acknowledges support from the Marsden Fund of the Royal Society of New Zealand.

This work was supported by the high-performance computing clusters Seondeok at the Korea Astronomy and Space Science Institute. This research made substantial use of \textsc{class} \citep{Blas:2011rf}, the \textsc{ChainConsumer} package \citep{samuel_hinton_2020_4280904}, the \textsc{Numpy} package \citep{book} and \textsc{Matplotlib} \citep{Hunter:2007}.

\section*{Data Availability}

No new primary data sets were generated in the production of this article. The Fisher matrix forecasts are available from the author upon request.



\bibliographystyle{mnras}
\bibliography{inflation_after_2030} 








\bsp	
\label{lastpage}
\end{document}